\documentclass{jhl}
\newenvironment{DataAvailability}{\subsection*{Data Availability}}{}
\definecolor{aa}{RGB}{46,111,165}
\usepackage[export]{adjustbox}
\usepackage{tikz}
\usepackage{xcolor}
\definecolor{aa}{RGB}{170,170,170}
\usepackage{fancyhdr}
\usepackage[most]{tcolorbox}

\title{Systematic Analysis of Ferroelectric Domain Dynamics in Periodically Poled Lithium Niobate Waveguides Using Two-Photon Microscopy and Digital Imaging Processing}

\author[psi,corr,cont]{Jiuyi Zhang}
\author[psi,cont]{Christopher Cullen}
\author[psi]{Matthew Konkol}
\author[psi]{Peng Yao}
\author[psi]{Timothy Creazzo}
\author[psi]{Janusz Murakowski}
\author[psi]{Ruidong Xue}
\author[psi]{Xiaofeng Zhu}
\author[ud]{Md Omar Faruk Rasel}
\author[ud]{Yash Kabra}
\author[psi,ud]{Dennis Prather}

\address[psi]{Phase Sensitive Innovations Inc, 125 Sandy Dr, Newark, DE, USA}
\address[ud]{Electrical and Computer Engineering Department, University of Delaware, 140 Evans Hall, Newark, DE, USA}

\thanks[cont]{These authors contributed equally to this work.}
\thanks[corr]{Correspondence to: Jiuyi Zhang: \url{zhang@phasesensitiveinc.com}}

\begin{abstract}
 We present a characterization and analysis methodology suitable for volume production for characterizing and optimizing x-cut thin-film periodically poled lithium niobate (PPLN) devices using two-photon (2P) microscopy with quantitative image processing. This method enables direct extraction of key structural parameters-such as duty cycle, phase-matching behavior, and domain uniformity-across large device sets in a non-destructive manner. By correlating 2P microscopy–derived structural metrics with systematic variations in poling conditions, we establish a scalable, image-driven approach for evaluating and improving PPLN fabrication. The resulting workflow supports wafer-level process development and accelerates the manufacturing and packaging of lithium-niobate photonic integrated circuits (PICs).
    \keywords{Periodically poled lithium niobate (PPLN), Two-photon microscopy, Ferroelectric domain inversion dynamics, Advanced Nano-fabrication, Advanced industrial micro- and nanoscale imaging technologies}
\end{abstract}

\begin{document}
\ExplSyntaxOn
\tl_gset:Nn \g__jhl_short_title_tl {Short Title}
\ExplSyntaxOff
\maketitle

\section*{Introduction}
\noindent

Thin-film lithium niobate (TFLN) has rapidly become a leading material platform due to its strong electro-optic response, large second-order nonlinearity, wide transparency window, and ability to support low-loss nanophotonic devices. These attributes enable a broad range of photonic technologies, including high-speed electro-optic modulators \cite{Wang2018,Zhang:21,Xu2020}, optical frequency combs \cite{Zhang2019,Wu2024, Song2025}, integrated lasers and optical amplifiers \cite{Chen:21,Shams-Ansari:22,10.1117/1.AP.5.3.034002}, nonlinear frequency conversion \cite{Qu2025, Park:22, Lu:19}, quantum photon-pair and squeezed-light sources \cite{PhysRevLett.124.163603,Shi2024,Harper:24}, spectroscopy and sensing \cite{Shams-Ansari2022,10.1063/5.0045869,Du2025}, and optical and neuromorphic computing systems \cite{Lin2024,PhysRevResearch.5.033206,Hu2025}.

Recent progress in wafer-scale periodically poled thin-film lithium niobate (PPLN) fabrication has significantly improved the accuracy and reproducibility of ferroelectric domain engineering. Advances such as real-time monitoring of poling signals \cite{Rao:19}, pre- and post-etch metrology for wavelength-accurate quasi-phase matching (QPM) design \cite{xin_wavelength-accurate_2025}, thermo-optic trimming modules \cite{Li:24}, adaptive compensation of waveguide-level nonuniformities \cite{Chen2024}, and micro-transfer printing of PPLN films onto large-area substrates \cite{Vande2025} have collectively enhanced yield and performance uniformity. Despite these advances, reliably achieving high-quality domain inversion-particularly for chirped, multi-period, or multi-waveguide devices requiring tightly controlled phase-matching wavelengths \cite{Huang2022,PhysRevLett.124.133904,Zhang2023}-remains challenging, with current fabrication workflows still exhibiting notable device-to-device variability.

\begin{figure*}[!t]
    \begin{adjustbox}{minipage=\dimexpr\linewidth-4pt\relax,
        rndframe={width=1pt,color=aa}{5pt}}
        \centering
        \includegraphics[width=\linewidth]{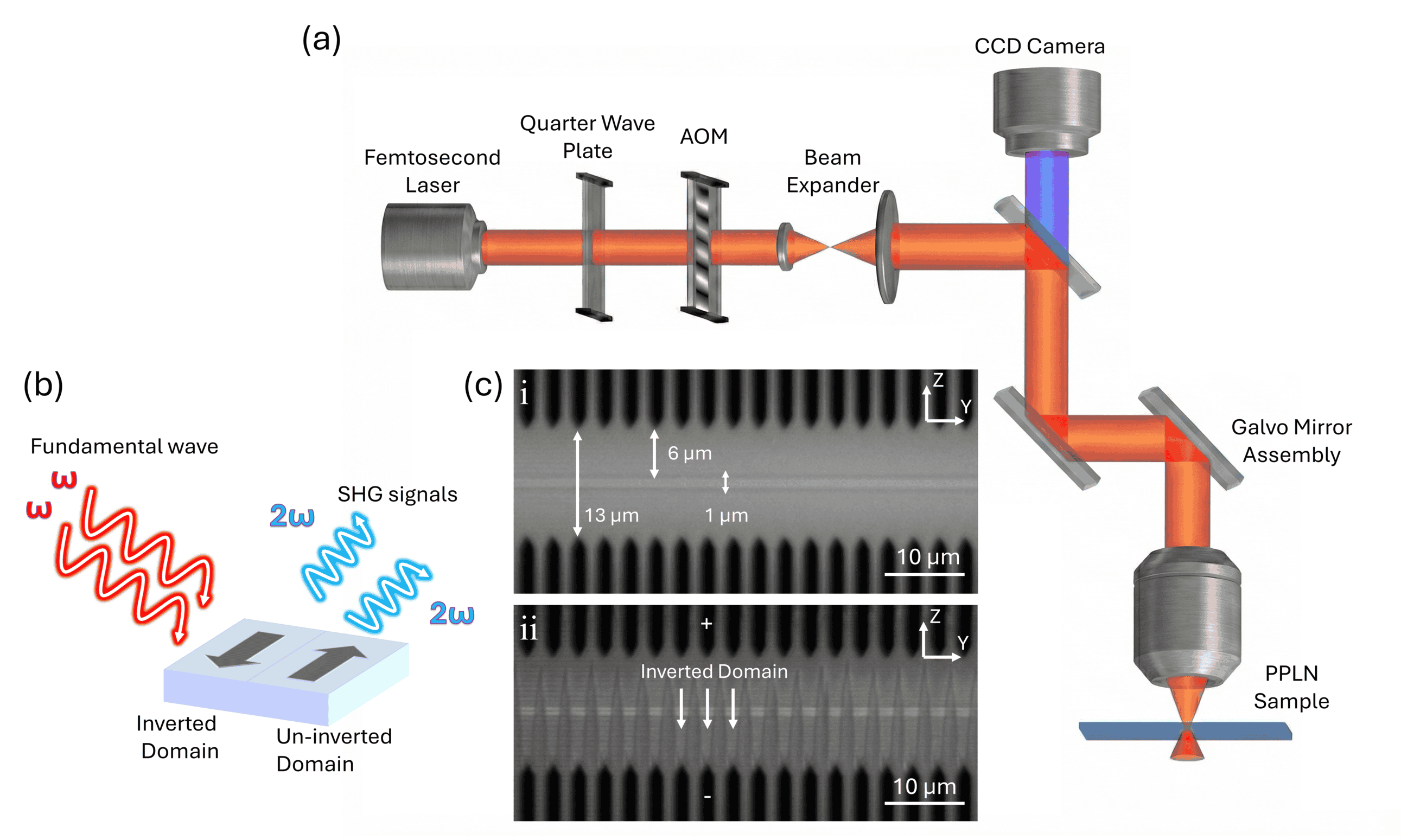}
        \caption{\textbf{Schematic of reflection non-interference two-photon (2P) microscopy for imaging periodically poled lithium niobate (PPLN) waveguides.}
        (a) In this configuration, two near-infrared excitation beams are tightly focused onto the PPLN waveguides on the chip, generating second-harmonic (SH) signals through localized two-photon absorption within both inverted and uninverted ferroelectric domains.
        (b) The SH waves generated in oppositely poled domains possess a $\pi$ phase difference, creating interference-enhanced contrast at domain boundaries and revealing variations in inversion depth.
        (c) Representative 2P images of (i) an unpoled TFLN waveguide and (ii) a periodically poled waveguide. The strong contrast between adjacent domains enables high-fidelity visualization of the poling pattern.}
        \label{fig:TPE}
    \end{adjustbox}
\end{figure*}

Moreover, several critical post-fabrication steps, including domain-pattern inspection of waveguides, quasi-phase matching spectral characterization, optical coupling, and thermo-electric packaging-remain slow, labor-intensive, and difficult to automate. The lack of a rapid, comprehensive assessment of poling quality, nonlinear spectral response, and packaging-induced variations therefore constitutes a major bottleneck for scaling TFLN photonic integrated circuits (PICs) toward high-volume manufacturing. As a result, the deployment of TFLN-based PICs in demanding applications such as quantum communication \cite{Tanzilli2002,PhysRevLett.134.230801} and quantum sensing \cite{Stokowski2023} is increasingly constrained by the availability of robust and high-throughput diagnostic tools.

Artificial intelligence (AI) and computer vision have transformed industrial inspection and quality-control workflows across many manufacturing sectors, enabling automatic anomaly detection, defect localization, and extraction of structural features. Similar techniques could greatly benefit integrated photonic manufacturing; however, their adoption for PPLN devices is constrained by a fundamental challenge: to obtain clear, high-resolution, wafer-scale images of ferroelectric domain structures after poling. Conventional characterization tools-such as piezoresponse force microscopy (PFM), Raman spectroscopy, and standard second-harmonic generation (SHG) microscopy-either lack the required throughput, depth sensitivity, or ability to distinguish partially inverted domains.

Industrial-scale integration of two-photon (2P) microscopy offers several critical advantages for ferroelectric domain imaging in TFLN. Originally developed for biological and tissue imaging, 2P microscopy has recently expanded into the integrated photonics domain through its application in two-photon lithography. As illustrated in Fig.~\ref{fig:TPE} (a), two near-infrared excitation beams are tightly focused onto the PPLN waveguide from opposite sides of the chip, generating second-harmonic (SH) signals through localized two-photon absorption within both inverted and uninverted ferroelectric domains. The resulting SH emission propagates downward and is collected in reflection by a high numerical-aperture (NA) objective. This imaging modality distinguishes fully and partially inverted domains with nanometer-scale spatial resolution and provides sub-wavelength sensitivity to variations in domain-penetration depth \cite{rusing_second_2019,doshi_multi-scale_2025}. As shown in Fig.~\ref{fig:TPE} (b), the intrinsic $\pi$-phase difference between adjacent oppositely poled domains produces strong interference-enhanced contrast at domain boundaries \cite{Sheng:10,cherifi-hertel_non-ising_2017,OEA-2024-0139AdityaDubey}, enabling high-fidelity visualization of complex domain patterns, an example of which is presented in Fig.~\ref{fig:TPE} (c). In addition to providing robust contrast, 2P microscopy is non-destructive, three-dimensional, and compatible with automated systems capable of fully programmable imaging workflows. Its suitability for real-time imaging further makes it advantageous for both pre- and post-packaging inspections, as well as rapid process feedback during wafer-scale fabrication \cite{BANYS2023170686}.

\begin{figure*}[!t]
    \begin{adjustbox}{minipage=\linewidth-4pt,rndframe={width=1pt,color=aa}{5pt}}
        \centering
        \includegraphics[width=0.85\linewidth]{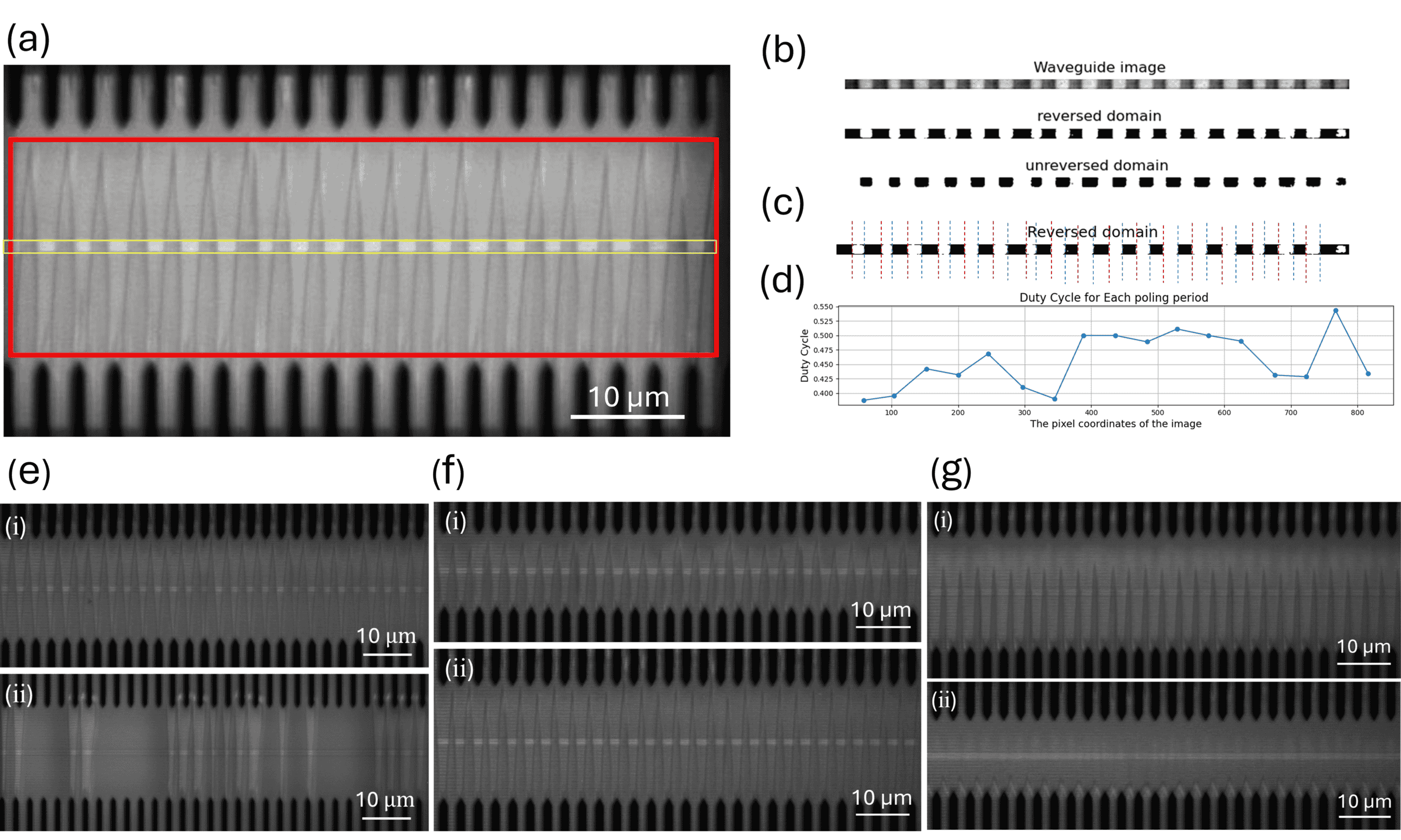}
        \caption{  \textbf{Workflow for SHG image processing and domain duty-cycle extraction, along with representative 2P microscopic images of under-poled, critically poled, and over-poled PPLN waveguides.}
        (a) A representative SHG microscopic image of a PPLN waveguide region, with the waveguide direction indicated by an arrow.
        (b) Binarized inverted-domain map obtained by thresholding, where bright regions correspond to poled (inverted) domains
        and dark regions correspond to unpoled (uninverted) areas.
        (c) Edge detection and length  analysis applied to the binarized image to identify the left (red dashed lines)
        and right (blue dashed lines) boundaries of each inverted domain.
        (d) Local duty-cycle distribution \( D_i = w_{\mathrm{rev},i}/w_{\mathrm{per},i} \) as a function of period index, showing the spatial variation and mean duty cycle across the analyzed waveguide.
        (e) Images of under-poled PPLN waveguides. (i) Example showing an inverted-domain duty cycle smaller than 50$\%$, indicating insufficient domain broadening. (ii) Example of an under-poled PPLN region exhibiting incomplete domain inversion.
        (f) Images of critically poled PPLN waveguides. (i) and (ii) show representative examples of PPLN waveguides with small and large electrode gaps, respectively, both exhibiting near-50$\%$ domain duty cycles characteristic of optimal poling.
        (g) Images of over-poled PPLN waveguides. (i) An example in which the inverted-domain width significantly exceeds the uninverted-domain width, indicating an over-poled condition. (ii) If the poling conditions that promote domain inversion are further intensified, the PPLN becomes fully inverted across the entire waveguide, as illustrated here.
        }
        \label{fig:image_analysis}
    \end{adjustbox}
\end{figure*}

In this work, we develop ferroelectric-domain inspection and analysis methods based on 2P microscopy by integrating a commercial photonic wire bonding (PWB, Vanguard SONATA1000) platform with advanced digital image-processing algorithms. By quantitatively extracting high-resolution structural descriptors of poled domains-most notably the local duty cycle-from 2P microscopic images, we systematically investigate how individual poling parameters and fabrication conditions influence the duty cycle and spatial uniformity of x-cut thin-film periodically poled lithium niobate (PPLN) waveguides, shows in Fig.~\ref{fig:image_analysis}. The parameter space analyzed spans a wide range of practical poling conditions, including poling voltage, electrode gap, electrode duty cycle, temperature, and pulse number, enabling a comprehensive, data-driven assessment of domain-quality trends across multiple fabrication runs.

Previous studies have extensively examined ferroelectric domain inversion in periodically poled x-cut thin-film lithium niobate, with particular emphasis on the roles of electric-field strength, pulse duration, and electrode geometry in governing domain nucleation, growth dynamics, and overall poling quality. These efforts have led to the identification of under- and over-poled morphologies and have guided empirical process optimization strategies\cite{Younesi:21}. In parallel, unified theoretical models that describe the dynamics of domain-evolution under applied electric fields have clarified the key stages of periodic poling\cite{Shur:2015}. More recently, in situ monitoring techniques have enabled real-time observation of the poling process in x-cut thin-film lithium-niobate photonic waveguides, culminating in wafer-scale domain inversion with high yield and improved uniformity\cite{Chen:2024}.

Building on these prior advances, the characterization and analysis methodology introduced here couples 2P microscopy with quantitative statistical and regression analysis to establish direct, image-derived correlations between poling conditions and domain-quality metrics. When further integrated with computational image analysis or AI-assisted computer-vision models, the proposed 2P microscopy-enabled workflow offers a scalable and high-throughput pathway for wafer-level process development, quality assurance, and advanced packaging of TFLN photonic integrated circuits.

\section*{Results}

\subsection*{Statistical Analysis of Duty-Cycle Variations in PPLN Waveguide Domain Structures}

To quantify how the measured duty cycle depends on the poling conditions, a multivariate regression analysis was performed using structural parameters extracted from two-photon microscopic images of PPLN waveguides. Five key fabrication variables were considered: electrode gap, electrode duty cycle, electric-field strength (V/$\mu$m), number of voltage pulses, and poling temperature. All continuous variables were standardized to zero mean and unit variance to enable direct comparison of the regression coefficients.

An extended regression model was constructed that incorporates linear, quadratic, and pairwise interaction terms to capture potential nonlinear dependencies between the poling parameters and the domain inversion behavior. The model is expressed as

\begin{equation}
y = \beta_{0} + \sum_i \beta_i x_i + \sum_i \beta_{ii} x_i^2 + \sum_{i<j} \beta_{ij} x_i x_j + \varepsilon,
\label{eq:duty_model}
\end{equation}
where $y$ denotes the standardized duty cycle measured from 2P microscopic images and $x_i$ represents the standardized predictors. This extended model provides a flexible analysis methodology for capturing both direct and coupled effects of fabrication variables.

\begin{table*}[t]
    \begin{adjustbox}{minipage=\linewidth-4pt,center,rndframe={width=1pt,color=aa}{5pt}}
        \centering
        \caption{Regression coefficients and significance summary for PPLN poling duty cycle (extended model with main, quadratic, and pairwise interaction terms). All predictors are standardized.}
        \label{tab:duty_cycle_regression_coefficients}
        \begin{tabular}{lcccc}
        \hline
        \textbf{Term} & \textbf{Coefficient ($\beta$)} & \textbf{$p$-value} & \textbf{Significance} & \textbf{Type} \\
        \hline
        Temperature ($temp$)            & $+0.52$  & $<10^{-5}$ & *** & Main Effect \\
        Number of Pulses ($pulses$)     & $+0.39$  & $<0.001$   & *** & Main Effect \\
        Field V/$\mu$m ($field$)        & $+0.21$  & $<0.05$    & *   & Main Effect \\
        Gap ($gap$)                     & $-0.28$  & $\approx 0.02$ & * & Main Effect \\
        Duty Cycle of Electrode ($elec\_duty$) & $+0.08$ & $>0.1$ & n.s. & Main Effect \\
        \hline
        $gap^2$                         & $+0.24$  & $\approx 0.04$ & *  & Quadratic \\
        $temp^2$                        & $-0.22$  & $\approx 0.03$ & *  & Quadratic \\
        $pulses^2$                      & $-0.18$  & $\approx 0.05$ & *  & Quadratic \\
        \hline
        $pulses \times temp$            & $+0.31$  & $<0.01$    & **  & Interaction \\
        $temp \times gap$               & $-0.19$  & $\approx 0.06$ & $\dagger$ & Interaction \\
        Others (non-significant)        & -       & $>0.1$     & n.s.& - \\
        \hline
        \end{tabular}

        \vspace{2mm}
        \noindent
        \textit{Note:}~“***” denotes $p<0.001$; “**” denotes $p<0.01$; “*” denotes $p<0.05$; “$\dagger$” denotes $0.05\le p<0.10$ (marginal); “n.s.” = not significant.
    \end{adjustbox}
\end{table*}

\begin{figure*}
    \begin{adjustbox}{minipage=\linewidth-4pt,rndframe={width=1pt,color=aa}{5pt}}
        \centering
        \includegraphics[width=0.85\linewidth]{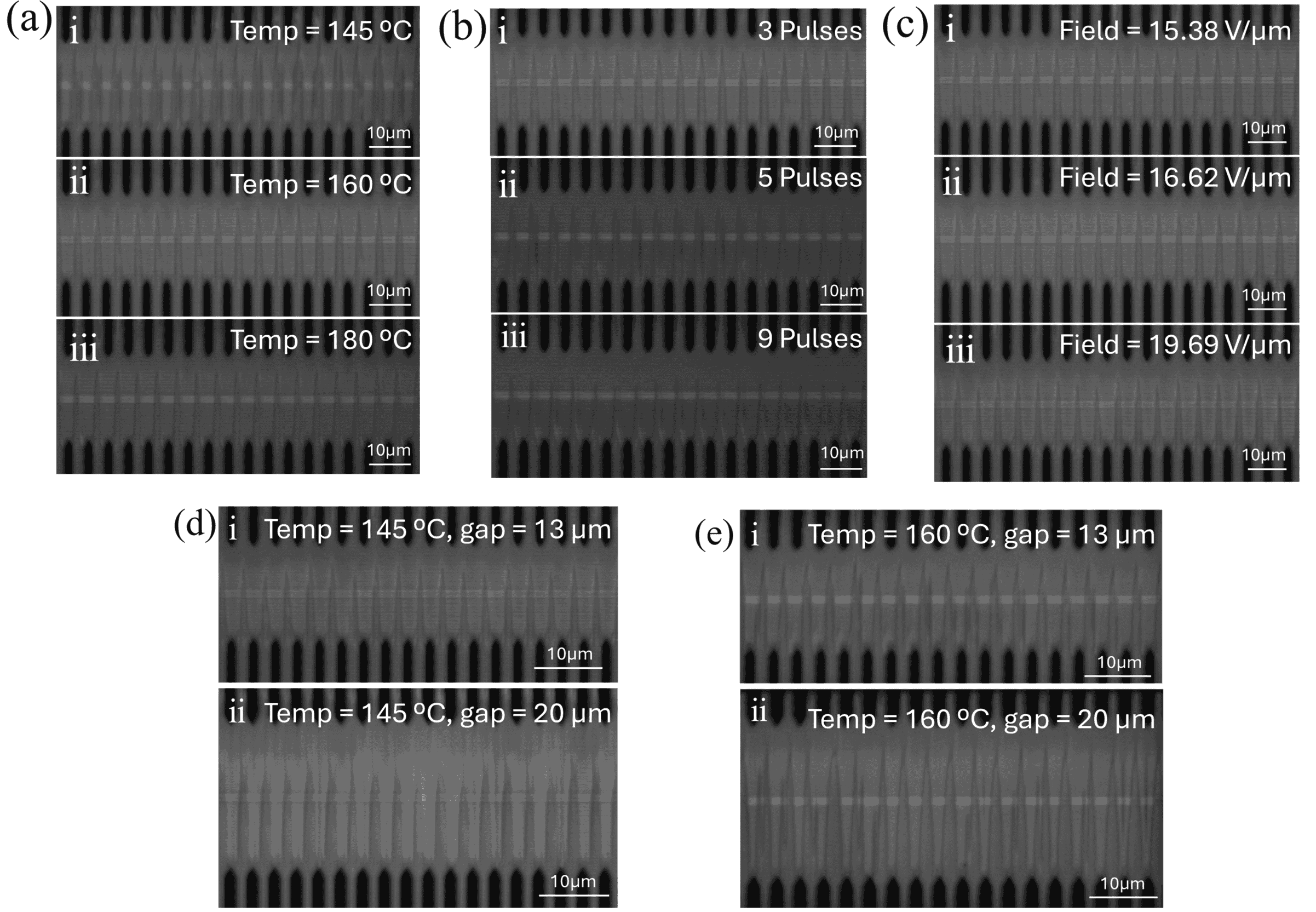}
        \caption{ \textbf{Second-harmonic generation (SHG) microscopy images showing the evolution of ferroelectric domain structures in PPLN waveguides under different poling conditions.}
\textbf{ (a). PPLN waveguides poled at different temperatures, showing the evolution of domain morphology with increasing poling temperature}. The SHG contrast highlights changes in domain uniformity and boundary sharpness as a function of the applied poling parameters.
Image i. corresponds to $145^{\circ}C$, Image ii. to $160^{\circ}C$, and Image iii. to $180^{\circ}C$.
All other poling parameters were kept constant: electric-field strength = 15.38 V/$\mu$m, number of pulses = 3, and electrode gap = 13 $\mu$m.
As the poling temperature increases, the fraction of inverted domains gradually increases, with the average duty cycles measured as 0.4195, 0.6357, and 0.6484 for Images i-iii, respectively.
\textbf{ (b). PPLN waveguides poled with different numbers of pulses, showing the evolution of domain morphology as the pulse count increases.}
Image i corresponds to 3 pulses, Image ii to 5 pulses, and Image iii to 9 pulses.
All other poling parameters were held constant: electric-field strength = 15.38 V/$\mu$m, temperature = $160^{\circ}C$, and electrode gap = 13 $\mu$m.
With increasing pulse number, the fraction of inverted domains increases progressively, and the average duty cycles are 0.6357, 0.6725, and 0.7503 for Images i-iii, respectively.
 \textbf{ (c). PPLN waveguides poled under different electric-field strengths, illustrating the effect of increasing field on domain morphology.}
Image i corresponds to a poling field of 15.38 V/$\mu$m, Image ii to 16.62 V/$\mu$m, and Image iii to 19.69 V/$\mu$m.
All other poling parameters were kept constant: number of pulses = 3, temperature = $160^{\circ}C$, and electrode gap = 13 $\mu$m.
As the poling field increases, the proportion of inverted domains grows gradually, with average duty cycles of 0.6145, 0.6357, and 0.7010 for Images i-iii, respectively.
 \textbf{ (d–e). Images comparing domain morphologies of PPLN waveguides fabricated with different electrode gaps under two poling temperatures.}
(d). Poling temperature = $145^{\circ}C$: Image i corresponds to a 13 $\mu$m electrode gap and Image ii to a 20 $\mu$m gap, with average duty cycles of 0.509 and 0.522, respectively.
(e) Poling temperature = $160^{\circ}C$: Image i corresponds to a 13 $\mu$m gap and Image ii to a 20 $\mu$m gap, with average duty cycles of 0.700 and 0.569, respectively.
All other poling parameters were identical: electric-field strength = 19.69 V/$\mu$m and number of pulses = 3.
At the lower temperature ($145^{\circ}C$), a larger electrode gap results in a slightly higher average duty cycle, whereas at the higher temperature ($160^{\circ}C$), a smaller gap yields a higher duty cycle, indicating an interplay between temperature and gap size in determining domain inversion efficiency.
}\label{fig:TPE_duty_cycle}
    \end{adjustbox}
\end{figure*}

The regression analysis shows that the model achieves excellent explanatory power (adjusted $R^2 = 0.91$, F-test $p < 10^{-6}$), indicating that the selected parameters explain the vast majority of the experimentally observed variation in the duty-cycle (Table~\ref{tab:duty_cycle_regression_coefficients}). Among all predictors, temperature is the dominant factor: Higher poling temperatures improve the uniformity and depth of domain inversion, consistent with thermally enhanced mobility of charged defects/ions that mitigates domain-wall pinning and reduces spatial non-uniformity in the effective switching (coercive) field\cite{10.1063/5.0234913}. The negative quadratic temperature term further indicates saturation at the upper end of the tested range. As shown in Fig.~\ref{fig:TPE_duty_cycle} (a) i-iii, increasing the poling temperature from $145^{\circ}C$ to $160^{\circ}C$ increases the duty cycle from 0.4195 to 0.6357, while a further increase to $180^{\circ}C$ results in only a marginal change to 0.6484, confirming that the temperature-driven improvement begins to plateau.

The \textbf{number of voltage pulses} also plays a substantial role. Increasing the pulse count promotes more complete inversion and leads to further broadening of the inverted domains, as shown in Fig.~\ref{fig:TPE_duty_cycle} (b). A weak negative quadratic dependence indicates diminishing returns beyond approximately six to seven pulses. The applied electric-field strength exhibits a moderate positive effect, reinforcing the role of the driving field in overcoming local coercive-field variations.

The electric-field strength (\textit{field V/$\mu$m}) contributes a moderate positive main effect ($\beta \approx \mathbf{+0.21}$, $p < 0.05$) and influences the duty cycle primarily through weak interactions with temperature and pulse number. As shown in Fig.~\ref{fig:TPE_duty_cycle} (c) i-iii, increasing the poling field from 15.38 V/$\mu$m to 16.62 V/$\mu$m and 19.69 V/$\mu$m raises the corresponding duty cycle from 0.6145 to 0.6357 and 0.7010. This trend indicates that the inverted domain width increases with stronger electric fields; however, the magnitude of this increase is modest compared to the effects of temperature and pulse number.

The \textbf{electrode gap} exhibits a nonlinear dependence: the linear term is negative, indicating improved inversion for narrower gaps because of stronger field localization. However, the positive quadratic term suggests that excessively narrow gaps may destabilize domain growth, producing a U-shaped overall trend. This implies that narrow gaps promote local field concentration and more complete inversion, while overly wide gaps lead to lateral field spreading and partial inversion, as shown in Fig.~\ref{fig:image_analysis} (e) ii.

Previous studies have shown that the duty cycle of the poling electrode pattern influences the lateral width of the inverted ferroelectric domains, thus affecting the resulting domain duty cycle in PPLN structures.\cite{kuznetsov2025wattlevelsecondharmonicgeneration,Chen:2024} However, within the design parameter space explored in this work, the electrode duty cycle exhibits the weakest direct contribution to the variation of the domain duty cycle, with a small main effect ($\beta \approx \mathbf{+0.08}$) that is not statistically significant ($p>0.1$). This result suggests that, relative to other poling parameters, the control of the domain duty cycle is less sensitive to variations in the geometry of the electrode pattern under the present operating conditions.

Several interaction terms are statistically significant, most notably the temperature–pulse interaction, which reveals a strong synergistic behavior: increasing the temperature enhances the effectiveness of multiple pulses by promoting thermally activated domain-wall motion and charge compensation. A marginal temperature-gap interaction is also observed, as shown in Fig.~\ref{fig:TPE_duty_cycle} (d)-(e). At lower temperature ($145^{\circ}$C), the PPLN with a 13 $\mu$m electrode gap exhibits a duty cycle of 0.509, compared to 0.522 for a 20 $\mu$m gap, indicating that smaller gaps initially yield slightly smaller duty cycles. However, at higher temperature ($160^{\circ}$C), duty cycles become 0.700 and 0.569 for the 13 $\mu$m and 20 $\mu$m gaps, respectively, demonstrating that temperature-induced improvements are substantially more effective for smaller electrode gaps. Other cross-terms (e.g., $field \times temp$, $field \times pulses$) are not statistically significant ($p>0.1$).

\begin{figure}[htbp]
    \centering
    \includegraphics[width=\textwidth]{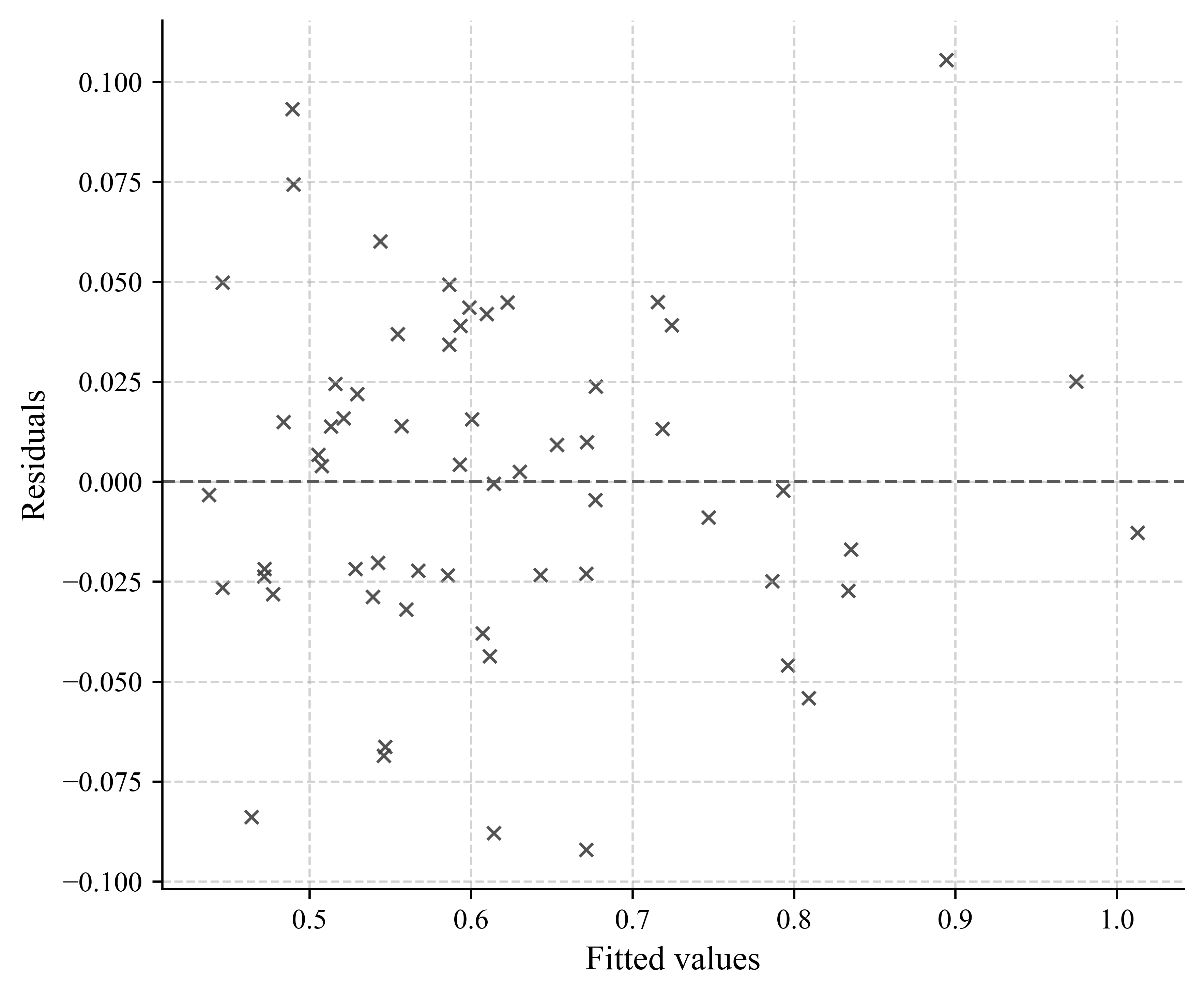}
    \caption{\textbf{Residuals vs Fitted - Extended Model.} Residuals scatter randomly around zero across the fitted range with no visible funneling, indicating no strong heteroscedasticity or leverage-driven patterns.}
    \label{fig:resid_ext}
\end{figure}

\begin{figure}[!t]
    \centering
    \includegraphics[width=\textwidth]{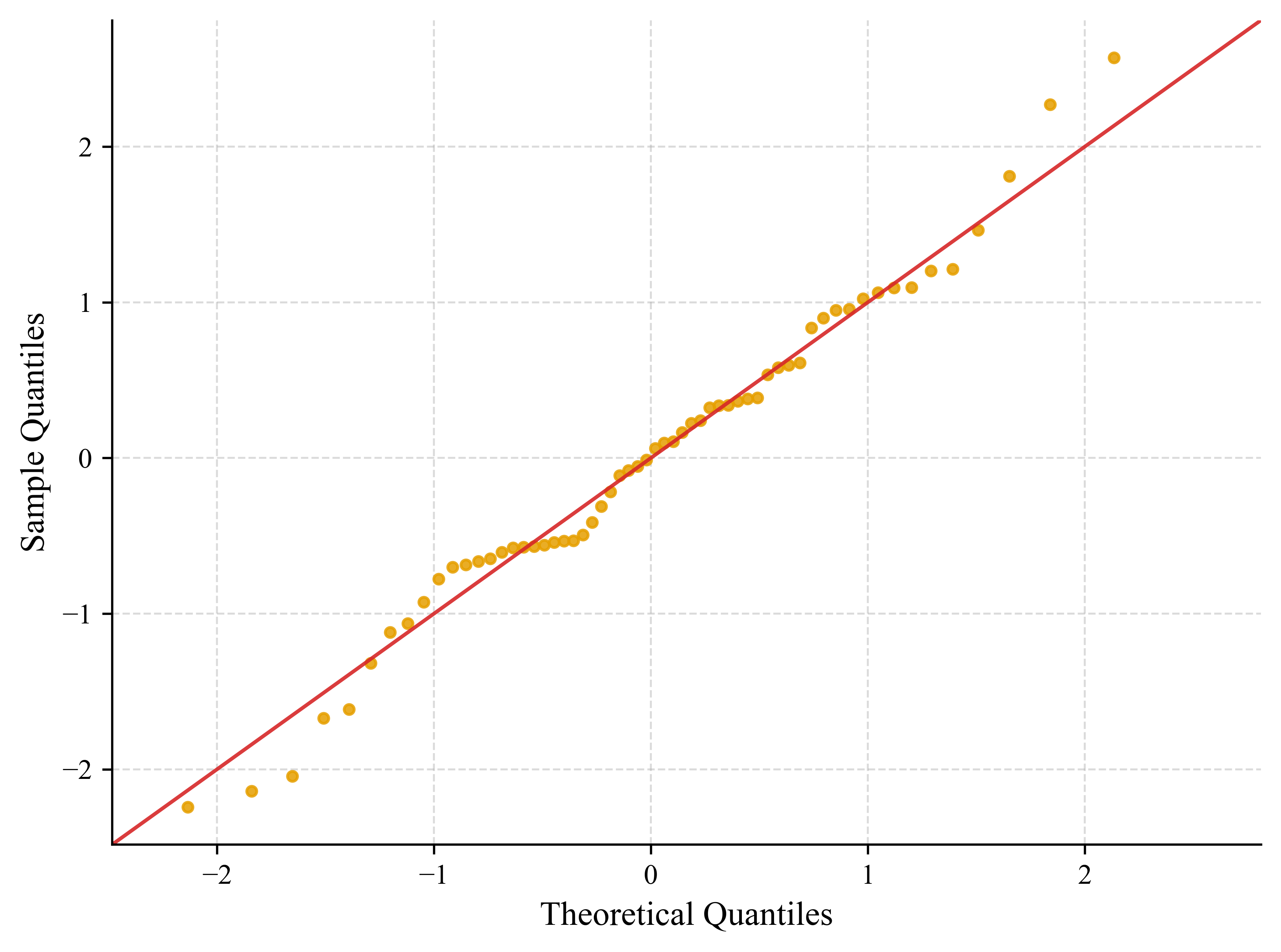}
    \caption{\textbf{Q-Q plot of extended-model residuals.} Points follow the 45$^\circ$ line closely with only minor tail deviations, consistent with approximate normality.}
    \label{fig:qq_ext}
\end{figure}

To validate the adequacy of the extended regression model, we performed residual diagnostics following standard practices for ordinary least squares (OLS) analysis~\cite{MontgomeryRegression}. Residual fitted plots were examined to assess potential systematic trends or heteroskedasticity, as such patterns would indicate model misspecification or omitted nonlinear structure; as shown in Fig.~\ref{fig:resid_ext}, the residuals exhibit no discernible structure or systematic dependence on the fitted values, suggesting that the model captures the dominant relationships between the poling parameters and the measured duty cycle. We further evaluated residual normality using normal quantile-quantile (Q-Q) plots, which underpin statistical inference in OLS regression~\cite{WilkQQ}. As illustrated in Fig.~\ref{fig:qq_ext}, the residuals closely follow the reference line, indicating that deviations from normality are minimal. Together, these diagnostics support the validity of the OLS-based inference and confirm that the extended model provides a reliable and statistically sound description of the dependence of domain duty cycle on the poling parameters.

Based on the regression result, a practical poling window can be identified that favors stable, high-duty-cycle inversion: temperatures around $160^{\circ}\mathrm{C}$, pulse counts of 5-7, electric-field strengths near 15 V/$\mu$m and electrode gaps of approximately 14-17 $\mu$m. These findings provide a robust, quantitatively grounded basis for optimizing PPLN fabrication processes.

\begin{table*}[!t]
    \begin{adjustbox}{minipage=\linewidth-4pt,center,rndframe={width=1pt,color=aa}{5pt}}
        \centering
        \caption{Significant terms in the extended regression model for duty-cycle standard deviation (excluding electrode duty).}
        \label{tab:uniformity_coeffs_noelec}
        \begin{tabular}{lccc p{7.0cm}}
        \hline
        \textbf{Term} & \textbf{Coefficient ($\beta$)} & \textbf{$p$-value} & \textbf{Significance} & \textbf{Interpretation} \\
        \hline
        $ I(\mathrm{gap}_s^2)$ & $+0.0170$ & $5.4\times10^{-7}$ & *** & Excessively small or large gaps increase nonuniformity \\
        $\mathrm{temp}_s$ & $-0.0147$ & $7.1\times10^{-7}$ & *** & Higher temperature improves domain uniformity \\
        $\mathrm{temp}_s:\mathrm{pulses}_s$ & $-0.0073$ & $1.7\times10^{-3}$ & ** & Synergistic stabilization at high temperature and multiple pulses \\
        $\mathrm{temp}_s:\mathrm{field}_s$ & $-0.0063$ & $3.3\times10^{-2}$ & * & Enhanced uniformity under combined high field and temperature \\
        $ I(\mathrm{temp}_s^2)$ & $+0.0048$ & $7.0\times10^{-2}$ & * & Mild U-shaped trend; extremely high $T$ may slightly degrade uniformity \\
        $\mathrm{field}_s:\mathrm{gap}_s$ & $+0.0055$ & $8.3\times10^{-2}$ & n.s. & Field–gap coupling; stronger fields across large gaps cause local nonuniformity \\
        $\mathrm{gap}_s$ & $-0.0055$ & $9.0\times10^{-2}$ & n.s. & Slight reduction in variation with increasing gap \\
        \hline
        \end{tabular}

        \vspace{2mm}
        \noindent\textit{Note:}~“***” denotes $p<0.001$; “**” denotes $p<0.01$; “*” denotes $p<0.1$; “n.s.” = not significant.
    \end{adjustbox}
\end{table*}

\subsection*{Statistical Analysis of Poling Domain Uniformity}

Another critical metric for assessing the poling quality of PPLN waveguides is the spatial uniformity of the ferroelectric domain widths. Non-uniform domain patterns introduce longitudinal fluctuations in the effective phase mismatch experienced by the guided optical mode, causing phase errors to accumulate along the propagation direction and ultimately degrading the nonlinear conversion efficiency. In addition to analyzing the duty cycle, the influence of different poling parameters on the uniformity of ferroelectric-domain formation in PPLN waveguides is therefore examined.

Using the digital image-processing workflow described earlier, the width of every individual inverted-domain segment was extracted from each two-photon microscopic waveguide image, and the standard deviation of these widths was calculated as defined in Eq.~\ref{eq:standard deviation}. This standard deviation provides a direct, quantitative measure of domain-width uniformity for each device, enabling systematic comparisons across the full range of poling conditions.

To determine how the poling parameters govern these spatial fluctuations, a multivariate regression model was constructed using the same analytical methodology used in the duty-cycle study. The predictors include four key parameters: electric-field strength (V/$\mu$m), number of voltage pulses, poling temperature, and electrode gap. Unlike the duty-cycle regression model, the electrode-duty-cycle parameter was excluded because it primarily influences the mean duty cycle rather than its spatial variation and contributes negligibly to domain uniformity. Excluding this parameter avoids unnecessary model complexity and reduces the risk of overfitting. The same extended regression form as in Eq.~\ref{eq:duty_model} was used, including linear, quadratic, and pairwise interaction terms, where $y$ is the standardized standard deviation of the measured duty-cycle of 2p microscopy and $x_i$ represents the standardized predictors.

The performance of the model was assessed using the adjusted coefficient of $R^2$, and the F-test for overall statistical significance. The extended model achieved an adjusted $R^2 = 0.519$ with $p_\mathrm{F} = 3.1\times10^{-6}$, indicating statistically significant explanatory capability for the observed variations in domain uniformity. The residual variance can be attributed to fabrication-induced imperfections, imaging limitations when resolving features approaching a single domain period in 2P microscopy, and segmentation uncertainty arising from the image signal-to-noise ratio on digital extraction accuracy. Although the uniformity model exhibits a lower $R^2$ than the duty-cycle model, its ability to reproduce experimental trends is highly meaningful. The close agreement between 2P microscopic-observed domain morphologies and predicted behaviors reinforces that the model captures the dominant dependencies between domain uniformity and the four poling parameters.

\begin{figure*}[!t]
    \begin{adjustbox}{minipage=\linewidth-4pt,rndframe={width=1pt,color=aa}{5pt}}
        \centering
        \includegraphics[width=\linewidth]{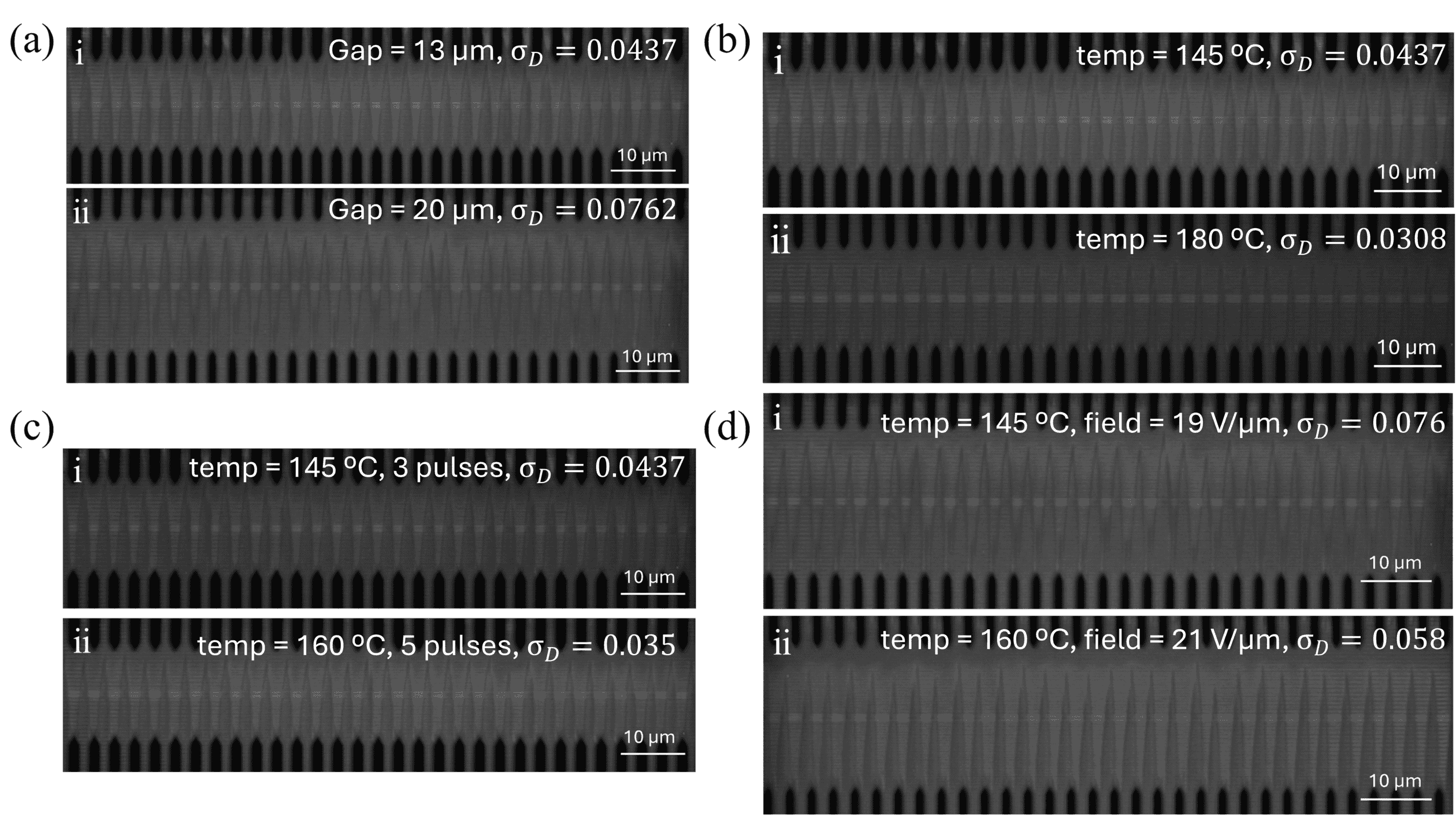}
        \caption{  \textbf {SHG microscopy images comparing the ferroelectric domain uniformity in PPLN waveguides under different poling conditions.}
 \textbf{(a). Comparison of domain uniformity under different poling electrode gaps at a fixed poling temperature of $145^{\circ}C$, applied field of 15.38 V/$\mu$m, and three voltage pulses.} (i) PPLN image with poling electrode gap = 13 $\mu$m, showing a standard deviation of duty cycle , \( \sigma_D \) = 0.0437; (ii) PPLN image with gap = 20 $\mu$m, exhibiting a larger \( \sigma_D \) = 0.0762. The comparison clearly indicates that increasing the poling electrode gap leads to a significant rise in domain uniformity fluctuations.
 \textbf{(b). Comparison of domain uniformity under different poling temperatures with a fixed poling electrode gap of 13 $\mu$m, applied field of 15.38 V/$\mu$m, and three voltage pulses. } (i) PPLN image at a poling temperature of $145^{\circ}C$, showing a \( \sigma_D \) = 0.0437; (ii) PPLN image at $180^{\circ}C$, with a lower \( \sigma_D \) = 0.0308. Across samples with temperatures increased from $145^{\circ}C$ to $160-180^{\circ}C$, the standard deviation of the duty cycle generally decreases by approximately 25–40 \%, clearly confirming the statistical conclusion that higher poling temperatures significantly improve domain uniformity.
\textbf{ (c). Synergistic improvement of domain stability under high temperature and multiple poling pulses with a fixed poling electrode gap of 13 $\mu$m.} (i) PPLN image at a poling temperature of $145^{\circ}C$ with three voltage pulses, showing a \( \sigma_D \) = 0.0437; (ii) PPLN image at $160^{\circ}C$ with five voltage pulses, exhibiting a reduced \( \sigma_D \) = 0.035. When high temperature and multiple pulses are applied simultaneously, the reduction in standard deviation is more pronounced than by increasing either parameter alone, indicating that thermally activated multiple-pulse poling promotes more complete domain-wall reversal and defect migration.
\textbf{ (d) Enhanced domain uniformity achieved through the combined effect of elevated temperature and stronger electric field with a fixed poling electrode gap of 20 $\mu$m.} (i) PPLN image at a poling temperature of $145^{\circ}C$ and field of 19.0 V/$\mu$m, showing a standard deviation of duty cycle = 0.076; (ii) PPLN image at $160^{\circ}C$ and 21.0 V/$\mu$m, exhibiting an improved standard deviation of duty cycle = 0.058. Simultaneous temperature increase and moderate field enhancement further improve poling uniformity. Elevated temperature not only reduces the coercive field but also facilitates charge injection and interfacial screening, resulting in a more homogeneous electric-field distribution during domain inversion.
}   \label{fig:TPE_duty_cycle_uniformity}
    \end{adjustbox}
\end{figure*}

The regression coefficients summarized in Table~\ref{tab:uniformity_coeffs_noelec} show that \textbf{temperature is the dominant factor} that governs the uniformity of the domain. The strong negative main effect ($\beta = -0.0147$, $p < 10^{-6}$) indicates that higher poling temperatures substantially suppress spatial fluctuations, yielding more homogeneous domain patterns. As shown in Fig.~\ref{fig:TPE_duty_cycle_uniformity} (b), increasing the poling temperature from $145^{\circ}\mathrm{C}$ to $180^{\circ}\mathrm{C}$ reduces $\sigma_D$ by approximately $40\%$, from 0.0437 to 0.035. This trend is consistent with previous ferroelectric studies that demonstrate that elevated temperatures lower the coercive field of LN, accelerate the motion of the domain-wall, and improve its smoothness and uniformity~\cite{10.1063/5.0029619}. These effects arise from thermally activated increases in ionic mobility and enhanced defect-mediated charge compensation near the Curie temperature, both of which reduce local coercive-field variability and stabilize domain-wall propagation~\cite{OEA-2024-0139AdityaDubey,10.1063/5.0234913}.

A small but noticeable quadratic temperature term ($\beta = +0.0048$, $p \approx 0.07$) suggests a weak U-shaped dependence, indicating that excessively high temperatures may slightly deteriorate uniformity. One potential mechanism is that over-driven domain-wall motion or charge accumulation near the electrode-LN interface may lead to subtle spatial irregularities at elevated temperatures \cite{BHOWMICK2017140}.

The\textbf{ electrode gap} exhibits linear and quadratic contributions. The mildly negative linear term ($\beta = -0.0055$) suggests that slightly larger gaps reduce field crowding and marginally improve uniformity, as Fig.~\ref{fig:TPE_duty_cycle_uniformity} (a) shows. However, the strong positive quadratic term ($\beta = +0.0170$, $p < 10^{-6}$) indicates that both wide and narrow gaps increase spatial variability. This nonlinear behavior demonstrates that an intermediate gap range optimizes the balance between local field confinement and uniform charge transport during poling.

A significant interaction between \textbf{temperature and pulse number} ($\beta = -0.0073$, $p = 1.7\times10^{-3}$) reveals a strong synergistic effect: at elevated temperatures, applying multiple voltage pulses further enhances uniformity by reducing domain-wall pinning and improving defect mobility, as shown in Fig.~\ref{fig:TPE_duty_cycle_uniformity} (c). Similarly, the \textbf{temperature-field} interaction ($\beta = -0.0063$, $p = 0.033$) indicates that higher temperatures enhance the effectiveness of the applied electric field, allowing more uniform switching even when the field strength is moderate. A weak but physically reasonable \textbf{field-gap} interaction ($\beta = +0.0055$, $p \approx 0.08$) suggests that strong electric fields applied across wide electrode gaps can distort the local field profile, causing slight domain-edge irregularities-an effect consistent with electrostatic modeling, as shown in Fig.~\ref{fig:TPE_duty_cycle_uniformity} (d).

Overall, the reduced four-parameter model is statistically robust. These results establish that \textbf{temperature is the primary process variable controlling domain uniformity}, followed by nonlinear electrode-gap effects and synergistic interactions involving temperature, pulse number, and electric field. Collectively, these insights provide quantitative guidelines for optimizing poling recipes: maintaining high but controlled poling temperatures, employing moderate electrode gaps, and applying multiple voltage pulses yield the most uniform domain inversion in TFLN.

\section*{Discussion}

The regression analysis presented in this work establishes a data-driven foundation for developing practical process-optimization tools. In future studies, these statistical models may be used as priors or baseline predictors and further integrated with machine-learning surrogate models, such as Gaussian processes or tree-based regressors, to predict optimal poling conditions for a broad range of device configurations, including different TFLN thicknesses, waveguide geometries, and poling periods. Building on such surrogate models, Bayesian optimization can be employed to efficiently search for optimal combinations of temperature, electric-field strength, pulse sequences, and electrode parameters under realistic fabrication constraints. In addition, integrating digital image analysis with artificial intelligence offers a complementary pathway, whereby AI models trained on the relationship between PPLN domain characteristics (e.g. duty cycle and domain uniformity) and normalized SHG efficiency could enable direct prediction of nonlinear conversion performance from PPLN images alone, eliminating the need for time-consuming nonlinear optical measurements on every individual device.

Overall, this data-driven optimization framework has the potential to support wafer-level process development, accelerate parameter tuning, and enhance reproducibility in the large-scale manufacturing of high-performance PPLN devices.

\section*{Materials and Methods}

\begin{figure*}[!t]
    \begin{adjustbox}{minipage=\linewidth-4pt,rndframe={width=1pt,color=aa}{5pt}}
        \centering
        \includegraphics[width=0.85\linewidth]{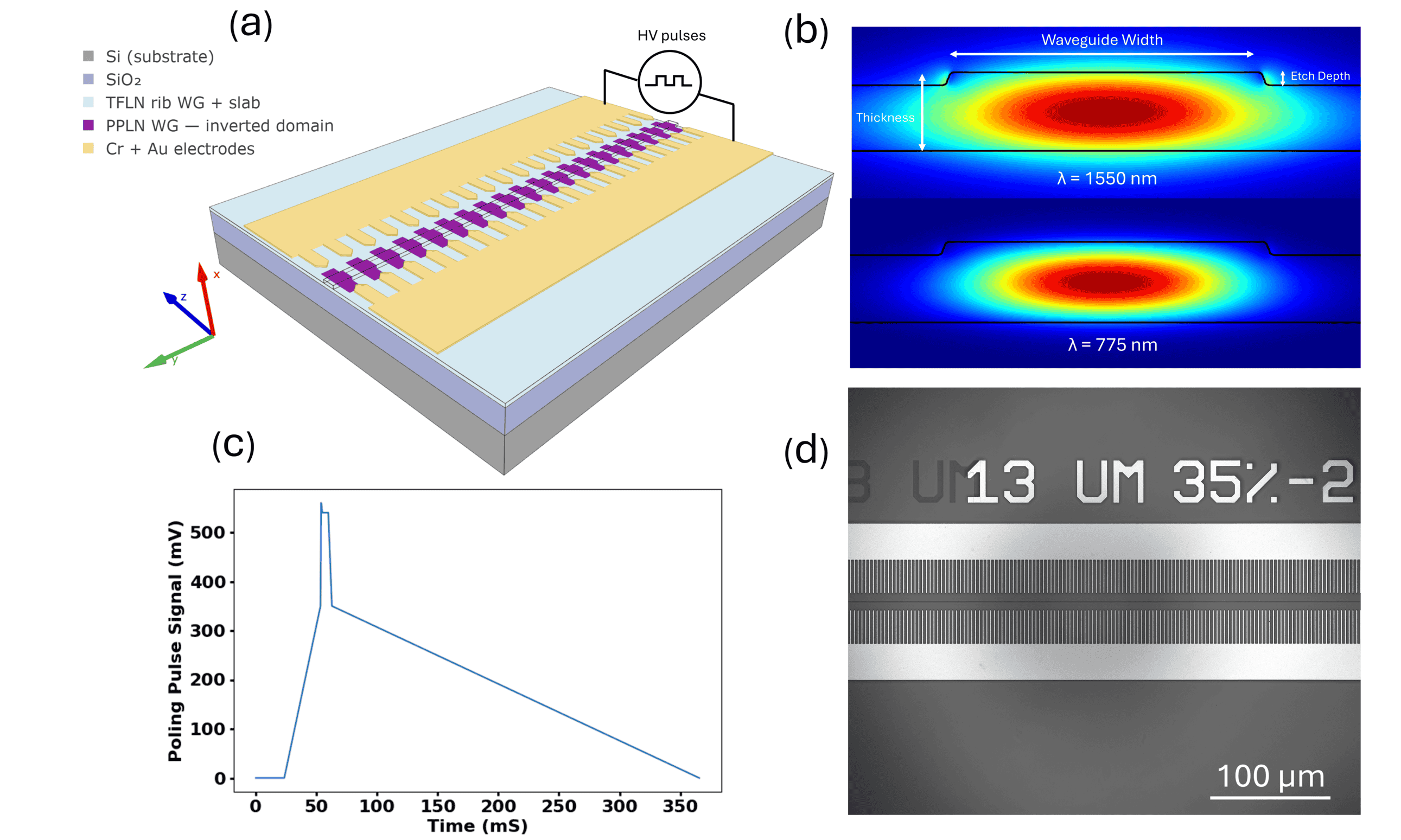}
        \caption{  \textbf{Schematic illustration of the PPLN waveguide with poling electrodes.}
        (a) Schematic of the ridge waveguide geometry. The structure consists of a 300-nm-thick lithium niobate thin film with a 50 nm etched rib, supported on a 3.7 $\mu$m SiO$_2$ buffer layer and a silicon substrate. A 300-nm-thick sputter SiO$_2$ protection layer was deposit on the top of PPLN sample. And a 170 nm Cr / 30 nm Au metal layer is deposited on top of the sputter SiO$_2$ layer as the poling electrode.
        (b) Mode profiles of the fundamental TE mode at the pump (1550 nm) and second-harmonic wavelengths (775 nm). The waveguide width is 1.2 $\mu$ m, and the rib waveguide is defined by a 50 nm shallow etch on a 300 nm x-cut LNOI platform.
        (c) Poling pulses waveform. High-voltage pulses are applied along the z-axis of the x-cut LNOI wafer.
        (d) Zoom-out 2P microscopic image of the PPLN sample with TFLN waveguide and poling electrode.}
        \label{fig:image_3D_illustration}
    \end{adjustbox}
\end{figure*}

\subsection*{Device fabrication}

x-cut TFLN waveguides were fabricated on a 300-nm device layer using electron-beam lithography (EBL, Raith EBPG5200ES) with a hydrogen silsesquioxane (HSQ) mask. The patterns were transferred into the LN layer using Ar ion milling (IntLVAC Nanoquest II IBE), yielding shallow-etched ridge waveguides with an etch depth of approximately 50 nm. Subsequently, a metal poling electrode stack consisting of 170 nm Cr and 30 nm Au was deposited on the chip surface, followed by a 300-nm sputtered SiO$_2$ cladding layer to improve electric-field confinement during poling, as shown in Fig.~\ref{fig:image_3D_illustration} (a). The PPLN structures were then formed using high-voltage pulses, with the pulse shape illustrated in Fig.~\ref{fig:image_3D_illustration} (c).

Compared with fully or deeply etched geometries, these shallow-ridge waveguides confine the optical mode primarily within the slab region, reducing sensitivity to sidewall roughness and minimizing propagation losses for both the pump and second-harmonic modes, as shown in Fig.~\ref{fig:image_3D_illustration} (b). Previous studies have also demonstrated that shallow-etched LN waveguides exhibit greater tolerance to fabrication-induced width variations~\cite{Zhao:20,PhysRevLett.124.163603}, making them well suited for isolating the influence of poling conditions on the formation of the PPLN domain and nonlinear optical performance.

\subsection*{Image analysis Methods}

Ferroelectric domain inversion was performed after the full waveguide and electrode patterning steps were completed, allowing systematic control and statistical evaluation across a large number of devices. A two-photon (2P) non-interference microscopy system, integrated into a commercial photonic wire bonding (PWB) tool (Vanguard SONATA1000), was used to characterize the resulting domain patterns. Due to destructive interference of the SHG generated in adjacent inverted and non-inverted regions, the 2P microscopic images exhibit high contrast at domain boundaries, enabling direct visualization of periodic ferroelectric structures with sub-micrometer spatial resolution \cite{OEA-2024-0139AdityaDubey}. From these images, key metrics such as local duty cycle, period length, and domain-wall morphology were extracted.

A two-stage digital image-processing workflow was developed to quantitatively evaluate the quality of the poling.

\textbf{Stage 1 - Binary domain-map generation:}
Each cropped 2P microscopic image corresponding to a single PPLN waveguide region was converted to an 8-bit grayscale format and binarized using Otsu's global thresholding method. The resulting binary mask delineated inverted (bright) and non-inverted (dark) domains. Vertical intensity projections were used to estimate the number of domains present and to provide initial estimates of the average poling period.

\textbf{Stage 2 - Domain-boundary extraction and duty-cycle computation:}

Edge detection was applied to the binarized images to accurately locate the lateral boundaries of each inverted domain. A run-length encoding (RLE) algorithm was then used to identify contiguous inverted regions and extract their start and end coordinates. From these pixel-level boundaries, both the period width $w_{\mathrm{per},i}$ and the inverted-domain width $w_{\mathrm{rev},i}$ were calculated for each domain. The local duty cycle was defined as

\begin{equation}
D_i = \frac{w_{\mathrm{rev},i}}{w_{\mathrm{per},i}}.\label{eq:duty cycle}
\end{equation}
The mean duty cycle $\bar{D}$ and its standard deviation $\sigma_D$ for each waveguide were evaluated as
\begin{equation}
\bar{D} = \frac{1}{N}\sum_{i=1}^{N} D_i, \qquad
\sigma_D = \sqrt{\frac{1}{N-1}\sum_{i=1}^{N} (D_i - \bar{D})^2},\label{eq:standard deviation}
\end{equation}
where $N$ is the number of periods detected. It is noted that the domain analysis was restricted to the poled regions confined within the $1 \mu\mathrm{m}$-wide PPLN waveguide. Within this narrow lateral window, the tapering of inverted-domain boundaries between adjacent electrodes typically introduces variations on the order of tens to a few hundred nanometers. Such variations are small compared to the waveguide width and do not significantly affect the extracted domain widths. Therefore, in this study, tapered or irregularly shaped lateral domain boundaries were treated equivalently without further sub-classification.

This workflow was applied to more than one hundred PPLN waveguides poled under systematically varied conditions including poling temperature, electric-field strength (V/$\mu$m), electrode gap, electrode duty cycle, and pulse number. The resulting data set provides a statistically robust basis for evaluating how each processing parameter influences the completeness, spatial periodicity, and overall uniformity of the domain inversion, allowing a comprehensive quantitative analysis of poling behavior in x-cut PPLN devices.

\begin{acknowledgement}
    \noindent
    The authors acknowledge the use of facilities and instrumentation supported by UDNF (UD Nanofabrication Facility) through the University of Delaware Materials Research Science and Engineering Center
\end{acknowledgement}

\begin{authorcontributions}
    \noindent
    J.Z designed the research framework, fabricated the device, performed the measurements, performed the data analysis, and wrote the manuscript. C.C fabricated the device, performed the measurement. M.K, P.Y, T.C ,J.M supervised the study, gave import technical advice. P.Y conceived the original idea, administered the project. M.K, R.X, X.Z participate the device fabrication. M.O.F.R, Y.K participate the measurement. All authors reviewed the manuscript.
\end{authorcontributions}

\begin{DataAvailability}
    \noindent
    All data are available from the corresponding authors upon reasonable request.
\end{DataAvailability}

\begin{conflictofinterest}
    \noindent
    The authors declare no competing interests.
\end{conflictofinterest}

\bibliography{ref}
\nocite{*}

\end{document}